\batchmode
\documentclass[twoside,12pt]{article}
\usepackage[T1]{fontenc}
\usepackage[latin9]{inputenc}
\usepackage{amsmath}
\usepackage{graphicx}
\usepackage{esint}

\makeatletter
\usepackage{epsfig}
\newcommand{\dw}{\mathrm{d}\omega}

\newcommand{\ddk}{\mathrm{d}^{3}k}
\newcommand{\dvk}{\mathrm{d}^{4}k}
\newcommand{\nb}{n_\text{B}}
\newcommand{\nf}{n_\text{F}}

\newcommand{\I}{\text{Im}}
\newcommand{\R}{\text{Re}}

\makeatletter\renewcommand{\@fnsymbol}[1]{\ensuremath{%
 \ifcase#1\or \dagger\or **\or {**}*\or   \mathsection\or \mathparagraph\or \|\or \star\or   \star\star\or {\star\star}\star \else\@ctrerr\fi}}\makeatother
\makeatother

\begin{document}

\title{\vspace{1cm}Equation of state for QCD matter in a quasiparticle
model}

\author{R.~Schulze\thanks{r.schulze@fzd.de}, B.~Kämpfer\\
\\
Forschungszentrum Dresden-Rossendorf, PF 510119, 01314 Dresden, Germany\\
Institut f\"ur Theoretische Physik, TU Dresden, 01062 Dresden, Germany}

\date{~\vspace{-1cm}}

\maketitle
\begin{abstract}
A phenomenological QCD quasiparticle model provides a means to map
lattice QCD results to regions relevant for a variety of heavy-ion
collision experiments at larger baryon density. We report on effects
of collectives modes and damping on the equation of state.
\end{abstract}
Strongly interacting matter is governed by the fundamental theory
of QCD, which can be solved numerically using Monte-Carlo calculations
on the lattice. However, reliable results are still limited to rather
small net baryon densities \cite{Eji06}. As an alternative approach
to obtain thermodynamic gross properties of the quark-gluon plasma,
a thermodynamic quasiparticle model (QPM) incorporating 1-loop QCD
in hard thermal loop (HTL) approximation can be utilized \cite{Pes00,BIR01,BKS07a,Sch08}.

Employing \cite{Sch08} the Cornwall-Jackiw-Tomboulis formalism, the
entropy density assumes the simple form of a sum $s=s_{g,\text{T}}+s_{g,\text{L}}+s_{q,\text{Pt.}}+s_{q,\text{Pl.}}+s'$
over partial entropy density contributions from four quasiparticle
excitations (transverse and longitudinal gluons, quarks and plasminos).
The residual interaction term $s'$ vanishes at 2-loop order for the
generating functional. Individual contributions read \begin{equation}
s_{i}\sim d_{i}\int_{\dvk()}\{\pi\varepsilon(\I D_{i}^{-1})\Theta\!\left(\xi_{i}\R D_{i}^{-1}\right)-\arctan\frac{\I\Pi_{i}}{\R D_{i}^{-1}}+\mbox{Re}D_{i}\mbox{Im}\Pi_{i}\Big\},\end{equation}
where $\int_{\dvk()}$ represents the convolution of the parentheses
$\{\}$ with the derivatives of the distribution functions with respect
to the temperature $T$, i.e.~$\int\ddk\int_{-\infty}^{\infty}\dw/(2\pi)^{4}(\partial\nb/\partial T)$
for the gluons and $\int\ddk\int_{0}^{\infty}\dw/(2\pi)^{4}(\partial\nf/\partial T+\partial\nf^{\text{A}}/\partial T)$
for quarks and plasminos (superscript A for antiparticles). The sign
constant $\xi_{i}$ is $-1$ for quasiparticles with real particle
interpretation (transverse gluons and quarks) and $+1$ for the collective
modes (longitudinal gluons and plasminos). $D_{i}$ ($\Pi_{i}$) stands
for the propagators (self-energies) of species $i$. From the entropy
density, the remaining state variables can be constructed in a self-consistent
manner.

To describe results of lattice QCD calculations at zero chemical potential,
a temperature shift is introduced into the running coupling $g^{2}$
changing it to an effective coupling $G^{2}$; the parameters of the
coupling (a scale parameter and the temperature shift) are then adjusted
to the lattice data.

To obtain the coupling $G^{2}$ at nonzero chemical potential $\mu$,
the self-consistency of the model and the stationarity of the thermodynamic
potential are employed, leading to a quasilinear partial differential
equation for the coupling (dubbed flow equation)\begin{equation}
a_{T}\frac{\partial G^{2}}{\partial T}+a_{\mu}\frac{\partial G^{2}}{\partial\mu}=b\end{equation}
with coefficients $a_{T,\mu}$ and $b$ listed in \cite{Sch08}. This
is the HTL QPM, as the HTL approximation is used for dispersion relations.

\begin{figure}
\noindent \begin{centering}
\hspace{-21px}\includegraphics[scale=0.9]{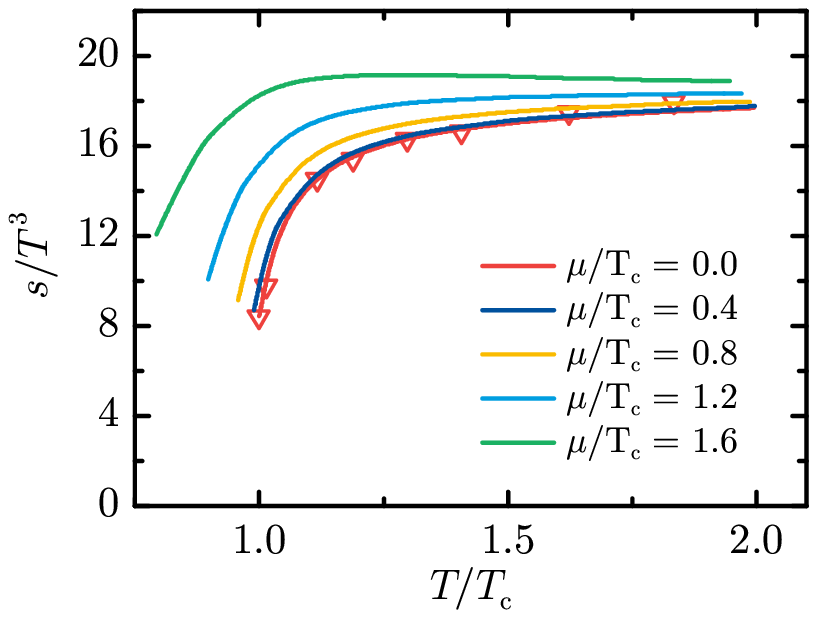}\quad\includegraphics[scale=0.9]{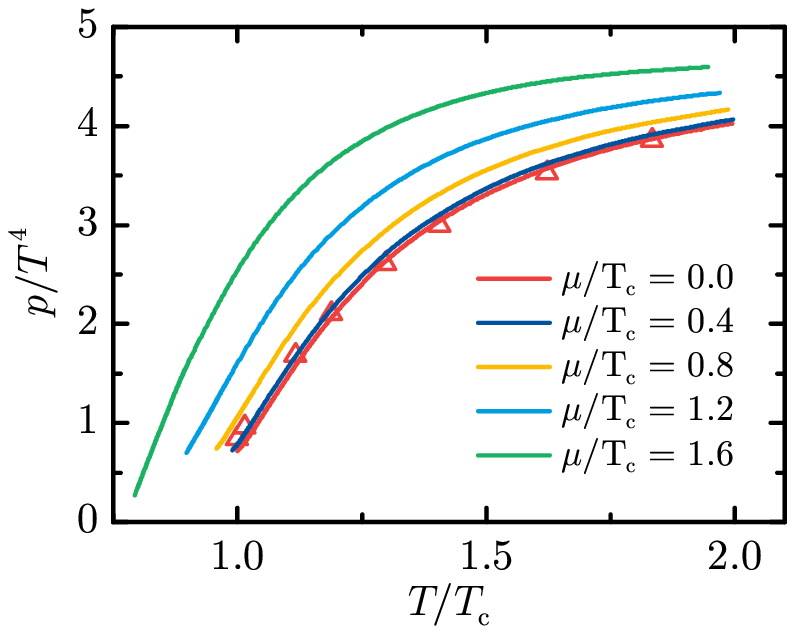}
\par\end{centering}

\caption{Scaled entropy density $s/T^{3}$ (left) and pressure $p/T^{4}$ (right)
of $2+1$ quark flavors as functions of the scaled temperature $T/T_{c}$
for several values of the quark chemical potential $\mu$. Lattice
data (symbols) for $\mu=0$ from \cite{Kar07}. The termination of
the curves at $T\leq T_{c}$ is at the conjectured transition line
to a confined state, cf.~\cite{Sch08,Sch07}.\label{fig:cuts}}

\end{figure}

For simplified versions of the HTL QPM, e.g.~neglecting collective
modes, the solution of the flow equation leads to ambiguities. It
was shown that collective modes and Landau damping as well as the
use of the momentum-dependent HTL dispersion relations are essential
to preserve the self-consistency of the model \cite{Sch07}. Utilizing
the full model, thermodynamic gross properties of the quark-gluon
plasma can be obtained. As an example, the entropy density and pressure
along lines of constant chemical potential are exhibited in Figure
\ref{fig:cuts}. From these state quantities, it is possible to provide
an equation of state for present and upcoming heavy-ion experiments
such as at RHIC, LHC \cite{BKS07b}, SPS and FAIR. In particular at
FAIR the baryon density effects covered by our model become severe.

One author (RS) thanks the organizers of the conference for support
and the opportunity to present his results.


\begin{thebibliography}{1}

\bibitem{Eji06}
S.~Ejiri, F.~Karsch, E.~Laermann, and C.~Schmidt, \emph{Phys. Rev.} D 73 (2006)
  054506

\bibitem{Pes00}
A.~Peshier, B.~K\"ampfer, and G.~Soff, \emph{Phys. Rev.} C 61 (2000) 045203

\bibitem{BIR01}
J.-P. Blaizot, E.~Iancu, and A.~Rebhan, \emph{Phys. Rev.} D 63 (2001) 065003

\bibitem{BKS07a}
M.~Bluhm, B.~K\"ampfer, R.~Schulze, and D.~Seipt, \emph{Eur. Phys. J.} C 49
  (2007) 205

\bibitem{Sch08}
R.~Schulze, M.~Bluhm, and B.~K\"ampfer, \emph{Eur. Phys. J.} ST 155 (2008) 177

\bibitem{Kar07}
F.~Karsch, \emph{J. Phys.} G 34 (2007) S627

\bibitem{Sch07}
R.~Schulze, \emph{Quasiparticle description of QCD thermodynamics: effects of
  finite widths, Landau damping and collective excitations}, Diploma thesis,
  Technical University Dresden (2007)

\bibitem{BKS07b}
M.~Bluhm, B.~K\"ampfer, R.~Schulze, D.~Seipt, and U.~Heinz, \emph{Phys. Rev.} C
  76 (2007) 034901

\end{thebibliography}
\end{document}